\documentclass{webofc}
\usepackage[varg]{txfonts}   
\newcommand{\ee}{e^{+}e^{-}}
\newcommand{\leplep}{\ell^{+}\ell^{-}}
\newcommand{\jp}{J/\psi}
\newcommand{\jpsi}{J/\psi}

\newcommand{\chiczp}{\chi_{c0}^{\prime}}
\newcommand{\chictwop}{\chi_{c2}^{\prime}}
\newcommand{\chictwo}{\chi_{c2}}

\newcommand{\pipi}{\pi^{+}\pi^{-}}

\newcommand{\piz}{\pi^{0}}

\newcommand{\ccbar}{c\bar{c}}

\newcommand{\rt}{\rightarrow}

\newcommand{\DDbar}{D\bar{D}}

\begin{document}
\title{Comment on the \boldmath{$X(3915)$} nonstandard hadron candidate}
%
%

\author{\firstname{Stephen Lars} \lastname{Olsen}\inst{1}\fnsep\thanks{\email{solsensnu@gmail.com}}
}

\institute{University of Chinese Academy of Science, Beijing 100049, CHINA} 

\abstract{
  I review the experimental evidence for the $X(3915)$, the candidate nonstandard meson
  associated with $\omega\jpsi$ resonance-like peaks in $B\rt K\omega\jpsi$ and
  $\gamma\gamma\rt\omega\jpsi$ near $M(\omega\jpsi)=3920$~MeV, and address the conjecture
  that it can be identified as the $\chictwop$, the radial excitation of the $\chictwo$
  charmonium state. Since the partial decay width for $B\rt K X(3915)$ is
  at least an order-of-magnitude higher than that for $B\rt K\chictwo$, its
  assignment as the $\chictwop$ is dubious. 
}
\maketitle
\section{Introduction}
\label{intro}
A number of meson candidates, dubbed the $XYZ$ mesons, that contain charmed-quark
anticharmed-quark ($\ccbar$) pairs but do not match expectations for any of the
unassigned levels of the $[\ccbar ]$ charmonium meson spectrum, have been observed in recent
experiments~\cite{Olsen:2017bmm}. In some cases, the distinction between the new
states that are nonstandard hadrons and conventional charmonium mesons remains
controversial.

\begin{figure}[htb]
\begin{minipage}[t]{70mm}  \includegraphics[height=0.75\textwidth,width=0.85\textwidth]{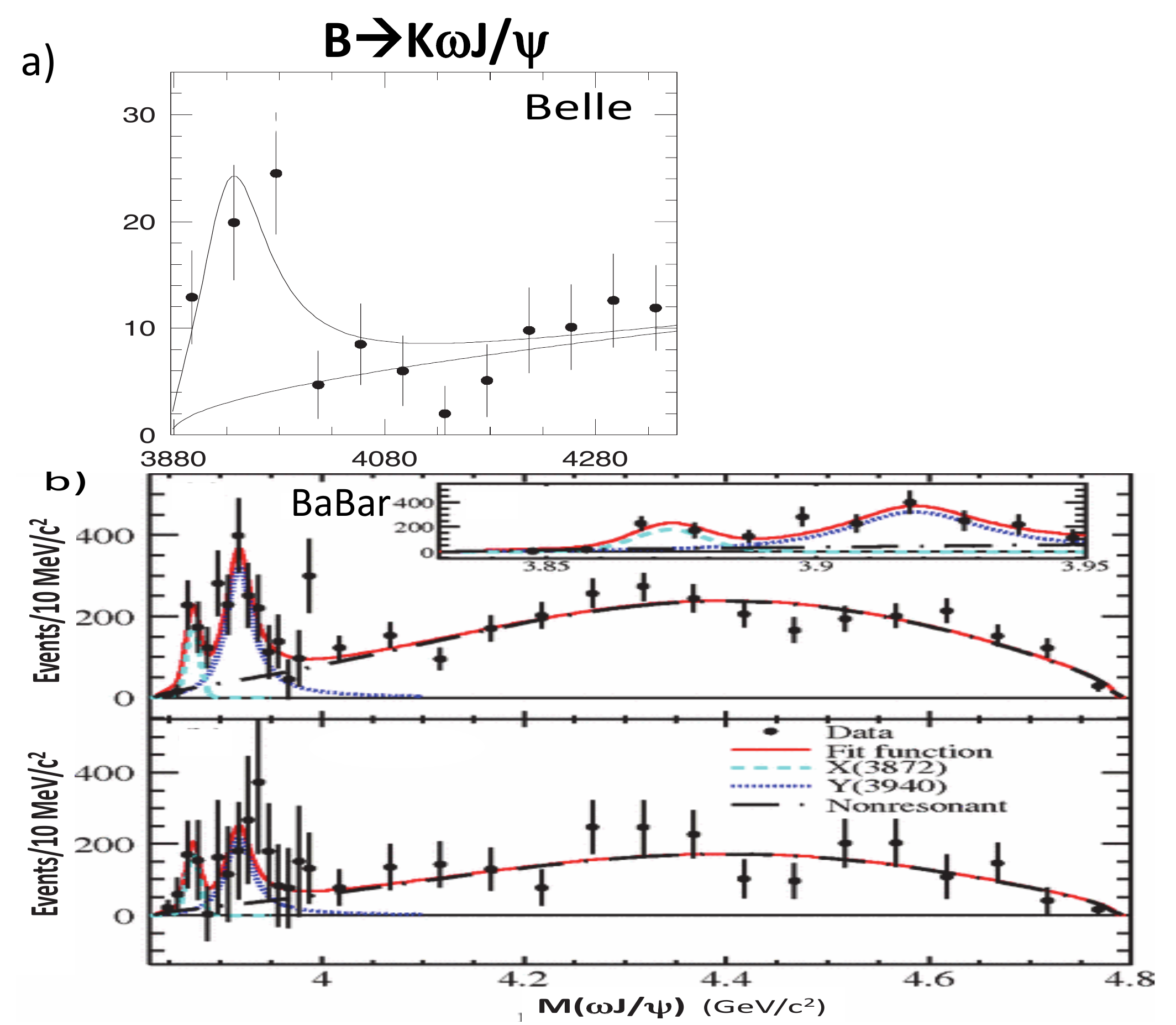}
\end{minipage}
\begin{minipage}[t]{70mm}
  \includegraphics[height=0.75\textwidth,width=0.80\textwidth]{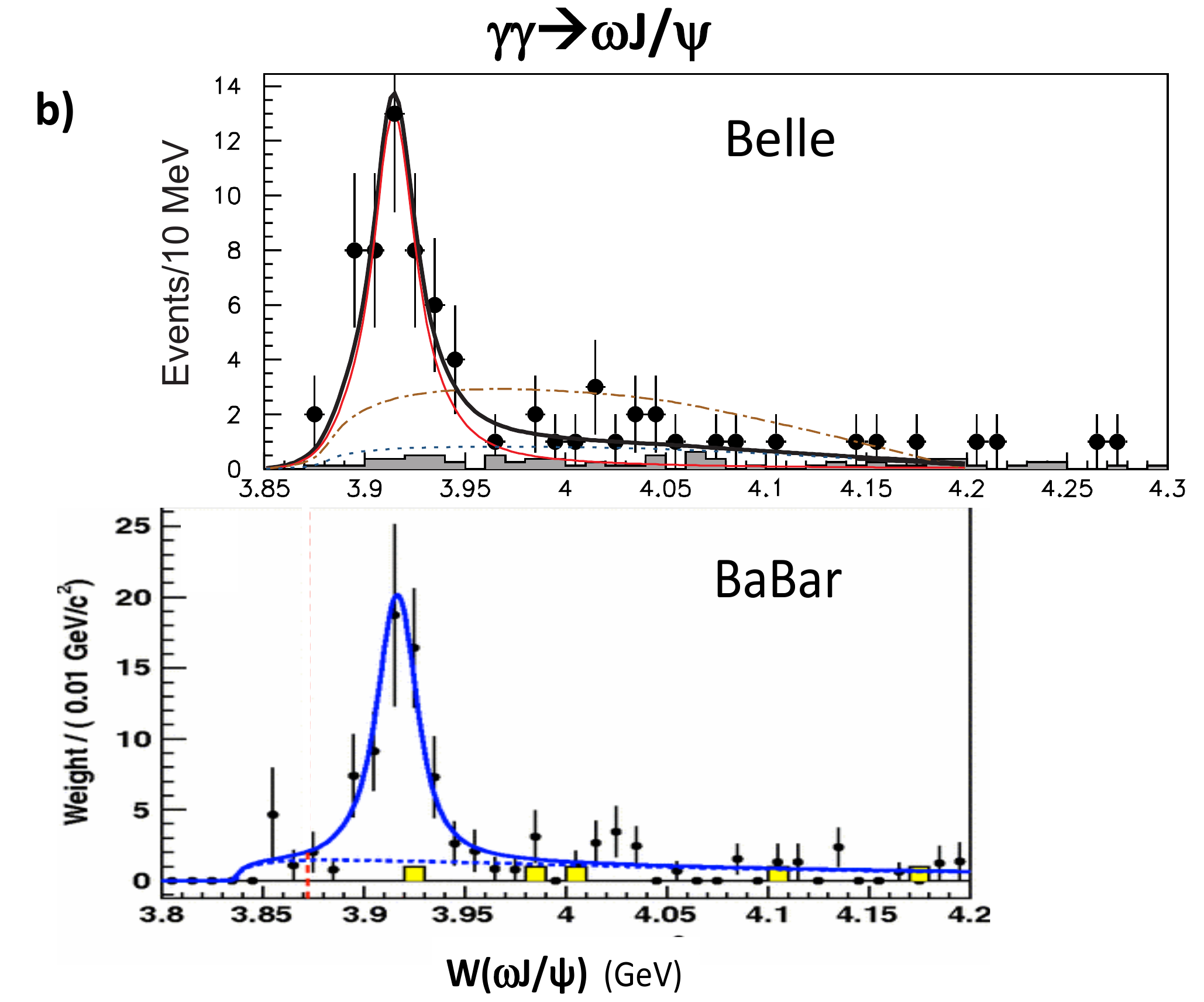}
\end{minipage}\hspace{\fill}
\caption{\footnotesize {\bf a)} The $\omega\jpsi$ invariant mass spectrum for $B\rt K\omega\jpsi$
  decays from {\it (top)} Belle~\cite{Abe:2004zs} and {\it (bottom)}
  Babar~\cite{delAmoSanchez:2010jr}. The low mass peak in the BaBar data is attributed to
  $X(3872)\rt\omega\jpsi$  (see inset); the higher mass peak is the
  $X(3915)\rt\omega\jpsi$ signal.  The Belle analysis did not consider the possible presence
  of an $X(3872)\rt\omega\jpsi$ signal. {\bf b)} The $\omega\jpsi$ mass spectrum for
  $\gamma\gamma\rt\omega\jpsi$ from {\it (top)} Belle~\cite{Uehara:2009tx} and {\it (bottom)}
  Babar~\cite{Lees:2012xs}.
}
\label{fig:fig1}
\end{figure}

This is especially the case for the $X(3915)$ that was first observed by
Belle~\cite{Abe:2004zs} and confirmed by BaBar~\cite{Aubert:2007vj,delAmoSanchez:2010jr}
as a near-threshold peak in the
$\omega\jpsi$ invariant mass distribution in exclusive $B\rt K\omega\jp$ decays (see
Fig.~\ref{fig:fig1}a). An
$\omega\jpsi$ mass peak with similar mass and width was seen in the two-photon fusion
process $\gamma\gamma\rt\omega\jp$, again by both Belle~\cite{Uehara:2009tx} and
BaBar~\cite{Lees:2012xs} (see Fig.~\ref{fig:fig1}b); BaBar reported its $J^{PC}$ to be
$0^{++}$. The similar masses and widths of
the peaks seen in the two production modes suggest that these are being produced
a single state (i.e., the $X(3915)$). The Particle Data Group's (PDG) average values for
the mass and width measurements from both production channels are~\cite{Tanabashi:2018oca}:
\begin{eqnarray}
     M(X(3915)) &=& 3918.4\pm 1.9~{\rm MeV}~~~{\rm and}~~~\Gamma(X(3915)) = 20.0 \pm 5.0~{\rm MeV},
     \label{eqn:x3915-m}
\end{eqnarray}
and the product branching fraction for $X(3915)$ production in $B^+$ meson decays is
\begin{eqnarray}
{\mathcal B}(B^+\rt K^+ X(3915))\times{\mathcal B}(X\rt\omega\jpsi)&=&3.0\pm 0.9\times 10^{-5}.
\label{eqn:prodbf}
\end{eqnarray}
The measured $\gamma\gamma\rt\omega\jpsi$ production rates are used to extract the
($J^{PC}$-dependent) widths:
\begin{eqnarray}
  \Gamma_{\gamma\gamma}(X(3915))\times {\mathcal B}(X\rt\omega\jpsi) &=&
  54\pm 9\ {\rm eV}~(0^{++})~~{\rm or}~~11.4\pm 2.7~{\rm eV}~(2^{++}).
  \label{eqn:gamee-wjpsi}
\end{eqnarray}

\section{The \boldmath{$X(3915)$} is not the \boldmath{$\chiczp$} charmonium state?}
\begin{figure}[h]
  \centering
  \includegraphics[height=0.25\textwidth,width=0.95\textwidth]{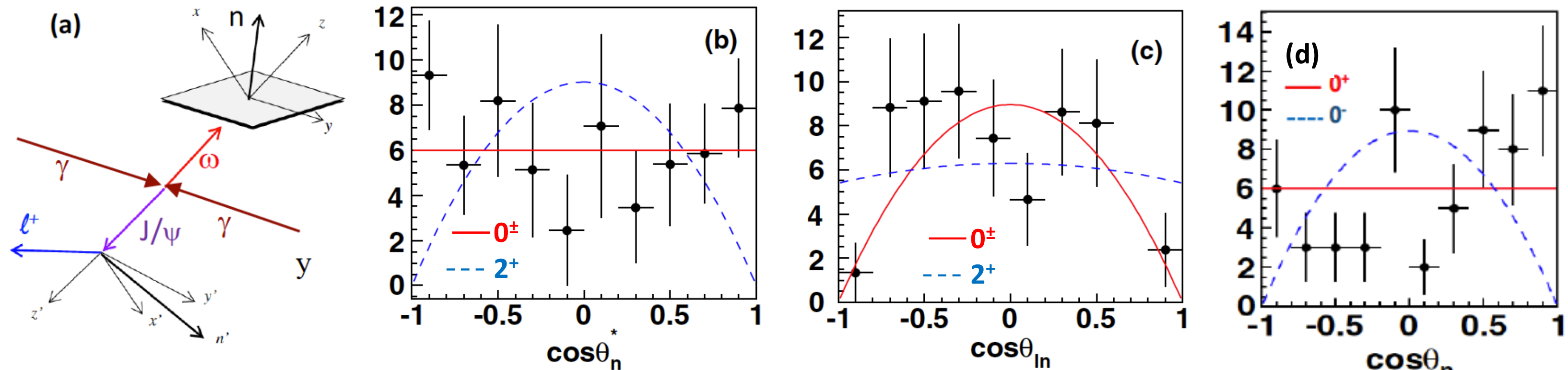}
  \caption{\footnotesize {\bf a)}~Directions used in the BaBar study of
    $\gamma\gamma\rt\omega\jp$,
    where $\omega\rt\pipi\piz$ and $\jp\rt\leplep$. {\bf b)}~Comparison of the $\cos\theta_{\rm n}^*$
    distribution with $J^{P}=0^{\pm}$ (solid red) and $2^+$ (dashed blue) expectations. {\bf c)}~The
    corresponding plot for $\cos\theta_{\rm ln}$.
    {\bf d)}~The $\cos\theta_{\rm n}$ distribution with expectations for $0^+$ in solid red
    and $0^-$ in dashed blue. (From ref.~\cite{Lees:2012xs}.)
 }
\label{fig:babar-angles}       
\end{figure}

The Babar group's $J^{PC}$ determination was based on an analysis of angular
correlations amongst the final-state particles in their
$\gamma\gamma\rt\omega\jpsi$ event sample~\cite{Lees:2012xs}. The important angles
for distinguishing $J=2^+$ from $J=0^{\pm}$
are $\theta_{\rm n}^*$, the angle between $\vec{\rm n}$, the normal to the $\omega\rt\pipi\piz$ 
decay plane, and the $\gamma\gamma$ axis in the omega rest frame, and  $\theta_{\rm ln}$, the angle
between $\vec{\rm n}$ and the direction of the $\ell^+$ from $\jpsi\rt\leplep$ decay (see
Fig.~\ref{fig:babar-angles}a).  Figure~\ref{fig:babar-angles}b shows the BaBar
$\cos\theta_{\rm n}^*$ distribution
together with the expectation for $J=0^{\pm}$ as a solid red line and $J=2^+$ as a dashed
blue curve.  There is a strong $\chi^2$ penalty for the near-zero event likelihood near
$\cos\theta_{\rm n}^* = \pm 1$  for the $J=2^+$ hypothesis to fluctuate {\it upward} to the
observed levels of $\sim 8$ and $\sim 9$ events, and this is the main support BaBar's
$J=0$ conclusion. The $J=2$ hypothesis seems to fit the BaBar $\cos\theta_{\rm ln}$
distribution (see Fig.~\ref{fig:babar-angles}c) better than
that for $J=0$.  But in this case, the likelihood of $\sim 6$ expected events near
$\cos\theta_{\rm ln}=\pm 1$ to fluctuate {\it downward} to the observed $\simeq 2$~events
is not so improbable. With $J=0$ established, the $0^+$ {\it vs.} $0^-$ discrimination
mostly relies on the angle $\theta_{\rm n}$,
which is the angle between the $\omega$'s flight path and $\vec{\rm n}$ in the $\omega\jpsi$
restframe.  The BaBar $\cos\theta_{\rm n}$ distribution shown in
Fig.~\ref{fig:babar-angles}d favors $0^+$ over $0^-$, mostly because of the $\simeq 10$
events near $\cos\theta_{\rm n}=+1$, where the $0^-$ expectation is zero.

BaBar's $J^{PC}=0^{++}$ assignment led them to suggest it as a suitable candidate
for the $2^3P_0$ charmonium state, commonly known as the $\chiczp$, and it was listed
as such in the 2014 PDG tables~\cite{Agashe:2014kda}. However, this assignment
had some problems and was challenged for a number of reasons~\cite{Guo:2012tv}:
the partial width for $X(3915)\rt\omega\jpsi$, which would be an OZI-suppressed
decay mode for a charmonium state, was too large; the lack of evidence for
$X(3915)\rt\DDbar$, which would be the dominant mode for the $\chiczp$; and
the small, $\simeq 9$~MeV, mass splitting between the $\chictwop$ and the
$X(3915)$, which is an order-of-magnitude lower than the smallest theoretical
estimates for $M_{\chictwop}-M_{\chiczp}$~\cite{Wang:2014voa,Olsen:2014maa}. This
assignment was finally put to rest in 2017 by Belle, when they reported the
observation of the $X^*(3860)$, a $\DDbar$ resonance with mass
$3862^{+47}_{-35}$~MeV in $\ee\rt\jpsi\DDbar$ annhilations with preferred
spin-parity of $0^{++}$~\cite{Chilikin:2017evr}.  These properties, particularly
the strong $\DDbar$ decay mode, match well the
expectations for the $\chiczp$, and the $X^*(3862)$ is clearly a much stronger 
candidate for this state than the $X(3915)$.

\section{Is it the \boldmath{$\chictwop$} charmonium state?}
\begin{figure}[htb]
    \centering
  \includegraphics[height=0.2\textwidth,width=0.81\textwidth]{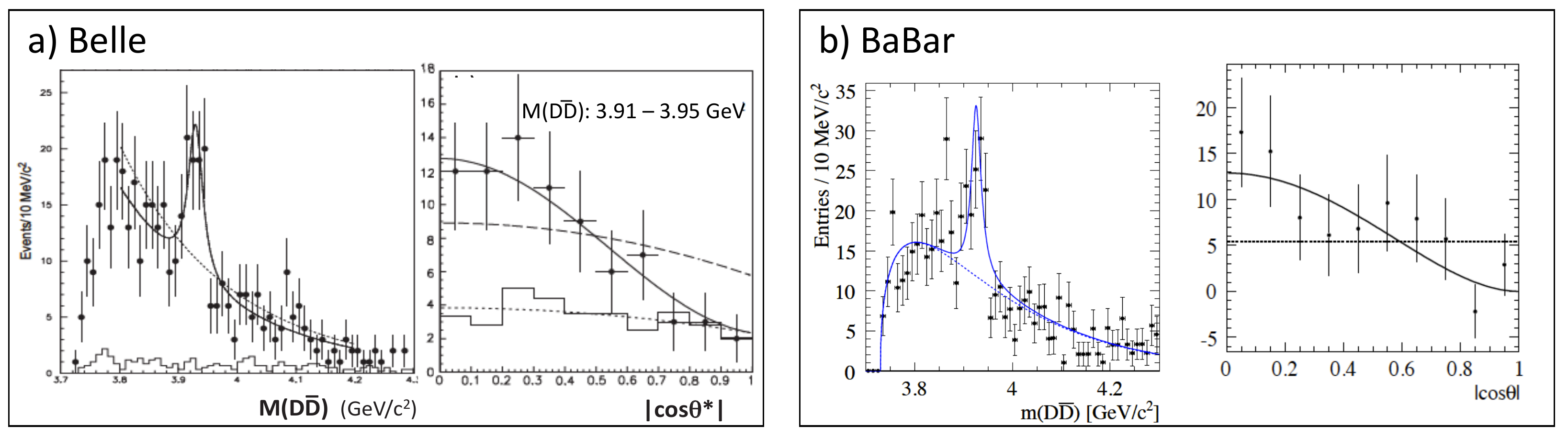}
\caption{\footnotesize {\bf a)} {\it left}: The $M(D \bar{D})$ distribution for 
  $\gamma\gamma\rt D\bar{D}$.  The open histogram the $D$ mass-sideband-determined
  background. The solid (dashed) curve
shows results of a fit that includes (excludes) a $\chi_{c2}^{\prime}$ signal.
{\it right}:~$dN/d|\cos\theta^*|$ for peak-region events with a
solid (dashed) curve showing $J=2$ ($J=0$) expectations.
The histogram is the non-resonant contribution. (From ref.~\cite{Uehara:2005qd}.)
{\bf b)}  Corresponding plots from BaBar~\cite{Aubert:2010ab}.
}
\label{fig:z3930}
\end{figure}

The $\chictwop$ was first spotted by Belle~\cite{Uehara:2005qd} and
subequently confirmed by BaBar~\cite{Aubert:2010ab} as a
prominent $M(D\bar{D})$ peak in the two-photon fusion process
$\gamma\gamma\rt D\bar{D}$ that has a distinct $\sin^4\theta^*$ production angle
dependence that is characteristic of a $J=2$ state. The
mass and width~\cite{Tanabashi:2018oca}:
\begin{eqnarray}
     M(\chi_{c2}^{\prime}) &=& 3927.2\pm 2.6~{\rm MeV}~~~{\rm and}~~~\Gamma(\chi_{c2}^{\prime}) = 24.0 \pm 6.0~{\rm MeV},
     \label{eqn:c2p-mass}
\end{eqnarray}
are consistent with charmonium expectations for the $\chi_{c2}^{\prime}$ and there are no
reasons to question this assignment. The Belle (BaBar) $M(D\bar{D})$ and $dN/d|\cos\theta^*|$
distributions are shown in Fig.~\ref{fig:z3930}a (b). Belle and BaBar measurements of its
two-photon production rate are also in good agreement and are characterized by the product
\begin{eqnarray}
  \Gamma_{\gamma\gamma}(\chictwop)\times {\mathcal B}(\chictwop\rt D\bar{D})
  &=& 210\pm 40~{\rm eV}.
\label{eqn:c2p-ggwidth}
\end{eqnarray}

BaBar's $J^{PC}=0^{++}$ assignment for the $X(3915)$
was based on a comparison to a $2^{++}$ scenario that only considered a
helicity-2 component ($h_2$) and ignored the possibility of any helicity-0 contribution.
This assumption of ``helicity-2 dominance'' originate from a theoretical analysis
that found that in two-photon production of tensor mesons, the helicity-0
component $(h_0)$ is zero in the non-relativistic limit~\cite{Krammer:1977an}. The
authors of ref.~\cite{Zhou:2015uva} point out that in the case of charmonium,
the suppression of helicity-0 contributions only applies to mesons that are
100\% $\ccbar$, which is generally considered to be unlikely for charmonium
mesons with masses above the $2m_D$ open-charm threshold (see, e.g.,
ref.~\cite{Pennington:2007xr}).

\begin{figure}[htb]
    \centering
  \includegraphics[height=0.2\textwidth,width=0.81\textwidth]{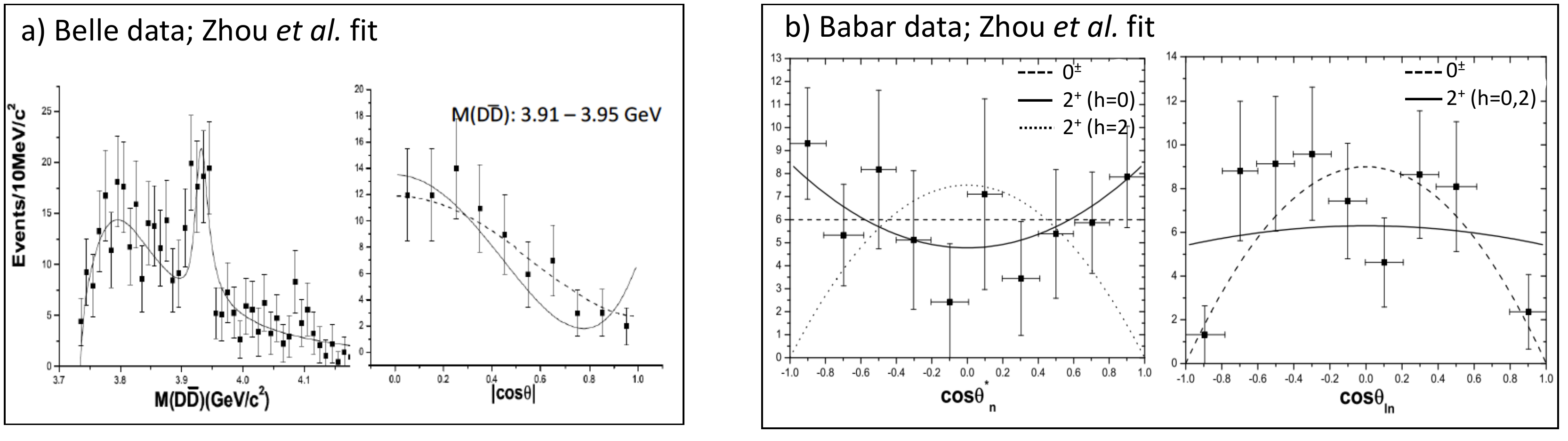}
  \caption{\footnotesize {\bf a)}  Belle $M(D \bar{D})$ ({\it left}) and
    $|\cos\theta^*|$ ({\it right}) distributions for
    $\gamma\gamma\rt D\bar{D}$ production.  The solid (dashed) curves
    show expectations for $h_0=0$  ($h_0=1.5 h_2$).  {\bf b)} BaBar $\cos\theta_{\rm n}^*$
    distribution ({\it left}) with a  solid (dotted) curve showing expectations for $2^{++}$ with
    $h=0$ ($h=2$); the dashed curve is for $0^{++}$. ({\it right}) The $\cos\theta_{\rm ln}$
    distribution with a solid curve for $2^{++}$ with $h=0\ {\rm or}\ 2$, and a
    dashed curve for $0^{++}$. (From ref.~\cite{Zhou:2015uva}.)
}
\label{fig:zhou-etal}
\end{figure}

This is important because if the $J^{PC}$ of the $X(3915)$ is $2^{++}$, the
mass peak identified with the $X(3915)$ could be conceivably be due to an
$\omega\jpsi$ decay mode of the $\chi_{c2}(2P)$ charmonium state.
The dashed lines in Fig.~\ref{fig:zhou-etal}a show the ref.~\cite{Zhou:2015uva}
comparison of the Belle $M(D\bar{D})$ and $|\cos\theta|$ with an $h_0\simeq 1.5 h_2$
mixture to represent the $X(3915)$. Figure~\ref{fig:zhou-etal}b) shows BaBar's
$\cos\theta_{\rm n}^*$ and $\cos\theta_{\rm ln}$ distributions with expectations
for $0^{++}$, and $2^{++}$ with $h=0$ \& $h=2$. With the inclusion of some $h=0$
contribution, the $\chi^2$ distinction between $0^{++}$ and $2^{++}$ angular distributions
is diminished and the authors conclude that the $X(3915)$ could be a $\chictwop$ state
that contains a sizable non-$\ccbar$ component.

\subsection{Other aspects of the \boldmath{$X(3915)=\chictwop$} assignment}

In addition to violating helicity-2 dominance, which ref.~\cite{Zhou:2015uva}
claims may not be a problem, there are other concerns with the $X(3915)=\chictwop$ assignment.
These are briefly  discussed here.

\subsubsection{Mass and width differences}
    Belle and BaBar  measurements of the $\gamma\gamma\rt\omega\jpsi$ mass peak,
    $3915\pm 4$ and $3919\pm 3$~MeV, respectively, are both lower, by $\simeq 2\sigma$,
    than their respective $\chictwop\rt D\bar{D}$ mass peak measurements, $3929\pm 5$ and
    $3927\pm 3$~MeV. Since the measurements reference
    well known masses -- $\omega$ and $\jpsi$ for the $X(3915)$ and 
    $D$-meson for the $\chictwop$-- systematic effects
    are small.

\begin{figure}[htb]
    \centering
  \includegraphics[height=0.225\textwidth,width=0.81\textwidth]{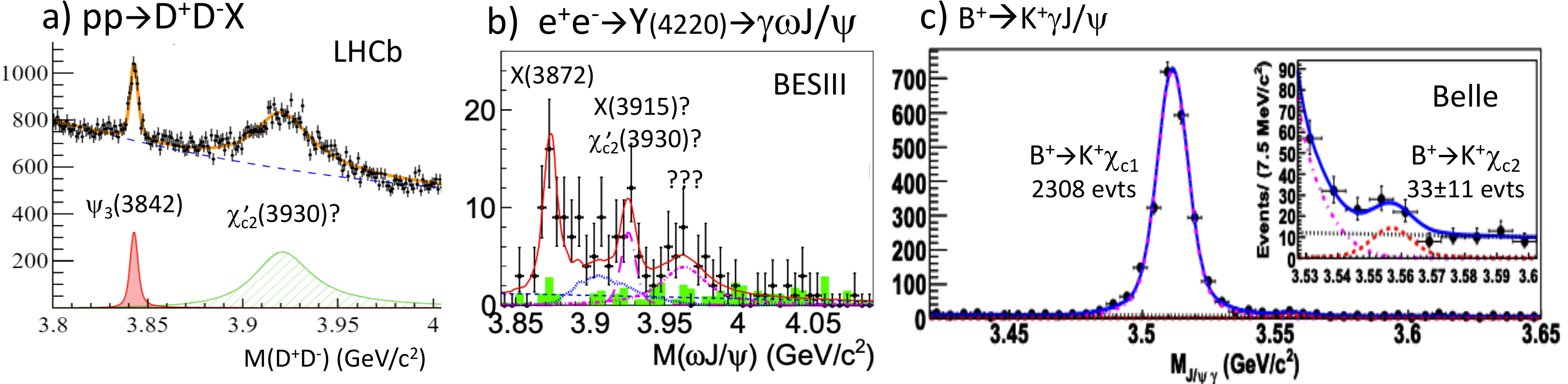}
  \caption{\footnotesize {\bf a)} The $M(D^+D^-)$ distribution for inclusive
    $D$-meson pair production at the LHCb. The peak at $3842$~MeV is the first
    observation of the $\psi_3$, the $1^3D_3$ charmonium level. The broader
    peak near $3920$~MeV is attributed by the LHCb group to the
    $\chictwop$~\cite{Aaij:2019evc}.
    {\bf b)} The  $M(\omega\jpsi)$ distribution for $\ee\rt Y(4220)\rt\omega\jpsi$
    events from BESIII. An $X(3872)\rt\omega\jpsi$ signal is evident. Additional
    peaks near 3925~MeV and 3960~MeV each have about $3\sigma$
    significance~\cite{Ablikim:2019zio}.
    {\bf c)}  $B^+\rt K^+\chi_{c1}$ and $K^+\chi_{c2}$ signals from the full Belle
    data set~\cite{Bhardwaj:2011dj}.
}
\label{fig:B2kchic2}
\end{figure}

    On the other hand, a recent LHCb report on the $M(D\bar{D})$ distribution for
    inclusive $D$-meson pair production in high energy proton-proton collisions
    included observation of a distinct peak in the $\chictwop$ mass region, shown in
    Fig.~\ref{fig:B2kchic2}a, with mass $M=3921.9\pm 0.6\pm 0.2$~MeV, $2\sigma$
    below the $\chictwop$ value listed in eqn.\ \ref{eqn:c2p-mass}~\cite{Aaij:2019evc}.
    The reported width, $\Gamma=36.6\pm 1.9\pm 0.9$~MeV, is $2\sigma$ higher than the
    eqn.\ \ref{eqn:c2p-mass} value. The LHCb group attributes this peak to the $\chictwop$.  

    Figure~\ref{fig:B2kchic2}b shows recent BESIII  $M(\omega\jpsi)$ results for
    $\ee\rt Y(4220)\rt\gamma\omega\jpsi$, where there is a
    strong $X(3872)\rt\omega\jpsi$ signal and $3\sigma$ ``evidence''
    for two higher mass peaks~\cite{Ablikim:2019zio}.  The fitted mass
    of the middle peak is $M=3926.4\pm 2.5$~MeV, near the Belle and BaBar
    results for $\chictwop\rt D\bar{D}$. Thus, the current situation with mass measurements
    is inconclusive.

    \subsubsection{A large OZI-violating {$\omega\jpsi$} decay width for a $[\ccbar ]$ meson}
    With the $\Gamma_{\gamma\gamma}\times {\mathcal B}$
    values listed in eqns.~\ref{eqn:gamee-wjpsi} and \ref{eqn:c2p-ggwidth},
    the $\chictwop$ assignment implies that
    \begin{eqnarray}
    \frac{{\mathcal B}(\chictwop\rt \omega\jpsi)}{{\mathcal B}( \chictwop\rt D\bar{D})}
      &=& 0.05\pm 0.02,
      \label{eqn:wjpsibf}
    \end{eqnarray}
    which is large for an
    OZI-rule-violating decay of an above-open-charm-threshold charmonium state,
    and more than an order-of-magnitude higher than the measured corresponding
    ratio for $\psi''\rt\pipi\jpsi$ and $D\bar{D}$. If $\chictwop\rt D\bar{D}$ and
    $D\bar{D}^*$ are the dominant decay modes and
    $\Gamma_{\chictwop}(D\bar{D}^*)\approx\Gamma_{\chictwop}(D\bar{D})$ (as predicted
    in ref.~\cite{Barnes:2005pb}), then
    $\Gamma_{\chictwop}(\omega\jpsi)>200$~keV (at the $\sim$90\% CL), and
    much larger than any measured OZI-violating width for a charmonium state.

  \subsubsection{${\mathcal B}(B\rt K\chictwop)>>{\mathcal B}(B\rt K\chictwo)$ ?}
    In 2011,
    with their full event sample accumulated over ten years, Belle
    reported $\sim 3\sigma$ evidence for $B^+\rt K^+\chictwo$ based on the
    $33\pm 11$~event signal shown in
    Fig.~\ref{fig:B2kchic2}c~\cite{Bhardwaj:2011dj}.
    The inferred branching fraction,
    ${\mathcal B}(B^+\rt K^+\chictwo)=1.1\pm 0.4\times 10^{-5}$, is smaller that the {\it product}
    branching fraction for $X(3915)\rt\omega\jpsi$ production in $B^+$ meson decays
    (eqn.\ \ref{eqn:prodbf}).
    Since ${\mathcal B}(\chictwop\rt D\bar{D})$ cannot exceed unity, eqn.\ \ref{eqn:wjpsibf}
    implies ${\mathcal B}(\chictwop\rt \omega\jpsi)<0.08$ (90\% CL). Thus, if the $X(3915)$
    produced in $B\rt K\omega\jpsi$ is the $\chictwop$, the $B$-meson decay width to $K^+\chictwop$
    would be more than an order of magnitude larger than that to $ K^+\chictwo$. This contradicts
    theoretical expectations that $B\rt K[\ccbar ]$ decay widths decrease with increasing
    radial $[\ccbar ]$ quantum numbers~\cite{Bodwin:1992qr}.

    Suppression of $B\rt K\chictwo^{(')}$ is not unexpected. The primary mechanism for
    $B$-meson ($\bar{b}q$) decays to $K[\ccbar]$ final states is $\bar{b}\rt \bar{c}$
    plus a virtual $W^+$ that, in turn, materializes as $c\bar{s}$. The final-state
    $c$- and  $\bar{c}$-quark form the $[\ccbar ]$ state and the $\bar{s}$- and
    ``spectator'' $q$-quark form the $K$.  This process is only allowed for
    $J^{PC}=0^{-+}, 1^{--}\ {\rm and}\ 1^{++}$ $[\ccbar ]$ states, decays to $[\ccbar]$
    states with other $J^{PC}$ values are higher-order and
    expected to be ``factorization suppressed''~\cite{Beneke:1999br}. The Belle
    results on $B\rt K\chictwo$ shown in Fig.~\ref{fig:B2kchic2}c demonstrate that for
    $J^{PC}=2^{++}$ $[\ccbar ]$ states, factorization suppression is very effective:
    ${\mathcal B}(B\rt K\chi_{c2} )< 0.04\times{\mathcal B}(B\rt K\chi_{c1})$ (90\% CL).

    \section{Summary and conclusions}

    Despite its observation by different experiments in a variety of production
    channels, the nature of the $X(3915)$ remains a mystery. If it is a nonstandard
    $XYZ$ meson, it cannot be easily interpreted by any of the proposed models for these states.
    For example: its mass is too low for a QCD-hybrid~\cite{Liu:2012ze}, and not near
    an appropriate threshold for a molecular state or a cusp effect~\cite{Olsen:2018ikz};
    the lack of evidence for a $\eta\eta_c$ decay mode~\cite{Vinokurova:2015txd} is
    problematic for a diquark-diantiquark assignmment~\cite{Lebed:2016yvr}. Thus, if it is
    an $XYZ$ meson, it is a very interesting one.

    The sum total of existing
    data on $\omega\jpsi$ and $D\bar{D}$ production in
    the $\sim 3925$~MeV mass region {\it cannot} be explained as
    being simply due to the $\chictwop$ charmonium state. While a (tenuous)
    case could be made that the near-3925~MeV mass peaks seen by the LHCb in
    $pp\rt D\bar{D} X$, Belle and BaBar in $\gamma\gamma\rt\omega\jpsi$ \& $D\bar{D}$
    and BESIII in $Y(4220)\rt\gamma\omega\jpsi$ are all due to decays of the $\chictwop$,
    the existing evidence is not conclusive. Moreover, a very strong case can be made
    {\it against} a $\chictwop$ interpretation of the $\omega\jpsi$ peak seen in
    $B\rt K\omega\jpsi$ decays.
    
    More refined  mass and width measurements are needed, and reliable, separate $J^{PC}$
    determinations for the $\omega\jpsi$ peaks produced via $\gamma\gamma$ fusion, radiative
    $Y(4220)$ transitions, and $B$-meson decays that eschew the
    helicity-2 dominance constraint are essential. The LHCb group has demonstrated that they can
    isolate clean $B^+\rt K^+\omega\jpsi$ signals with good efficiency~\cite{Andreassia:2014phd} and
    I look forward to high-statistics results from them in the near future.

    \section{Acknowledgements}
    I congratulate Phi-to-Psi-2018 organizers for an interesting and provocative meeting. This work
    is supported by the CAS President’s International Fellowship Initiative.
%
\bibliography{phipsi2018_olsen}
\end{document}